\documentclass[letterpaper]{article} 
\usepackage{aaai2026}  
\nocopyright
\usepackage{times}  
\usepackage{helvet}  
\usepackage{courier}  
\usepackage[hyphens]{url}  
\usepackage{graphicx} 
\usepackage{natbib}
\usepackage{caption}  
\usepackage{algorithm}
\usepackage{algorithmic}
\usepackage{newfloat}
\usepackage{listings}

\usepackage{color}
\usepackage{booktabs} 
\usepackage{amsfonts} 
\usepackage{amsmath}
\usepackage{pifont}
\usepackage{amssymb}

\DeclareCaptionStyle{ruled}{labelfont=normalfont,labelsep=colon,strut=off} 
\floatstyle{ruled}
\newfloat{listing}{tb}{lst}{}
\floatname{listing}{Listing}
\newcommand{\mymodel}{\textbf{\textit{FakeHunter}}~}
\newcommand{\mydataset}{\textbf{\textit{X-AVFake}}~}

\title{Memory-Anchored Multimodal Reasoning for Explainable Video Forensics}

\author{
Chen Chen\textsuperscript{\rm 1}\thanks{These authors contributed equally to this work.},
Runze Li\textsuperscript{\rm 2}\footnotemark[1],
Zejun Zhang\textsuperscript{\rm 3},
Pukun Zhao\textsuperscript{\rm 1},
Fanqing Zhou\textsuperscript{\rm 1},
Longxiang Wang\textsuperscript{\rm 4},
Haojian Huang\textsuperscript{\rm 5}\thanks{Corresponding author}
}

\affiliations{
\textsuperscript{\rm 1}Guangdong University of Finance and Economics \quad
\textsuperscript{\rm 2}Westlake University \quad
\textsuperscript{\rm 3}University of Southern California \quad
\textsuperscript{\rm 4}Chongqing University \quad
\textsuperscript{\rm 5}The University of Hong Kong\\
\texttt{Allen821@student.gdufe.edu.cn},\;
\texttt{lirunze@westlake.edu.cn},\;
\texttt{zejunzha@usc.edu},\;
\texttt{zhaopukun@student.gdufe.edu.cn},\;
\texttt{zhoyfanqing@gmail.com},\;
\texttt{longxiangwang@stu.cqu.edu.cn},\;
\texttt{haojianhuang@connect.hku.hk}
}

\begin{document}

\maketitle

\begin{abstract}
We address multimodal deepfake detection requiring both robustness and interpretability by proposing FakeHunter, a unified framework that combines memory guided retrieval, a structured Observation-Thought-Action reasoning loop, and adaptive forensic tool invocation. Visual representations from a Contrastive Language-Image Pretraining (CLIP) model and audio representations from a Contrastive Language-Audio Pretraining (CLAP) model retrieve semantically aligned authentic exemplars from a large scale memory, providing contextual anchors that guide iterative localization and explanation of suspected manipulations. Under low internal confidence the framework selectively triggers fine grained analyses such as spatial region zoom and mel spectrogram inspection to gather discriminative evidence instead of relying on opaque marginal scores. We also release X-AVFake, a comprehensive audio visual forgery benchmark with fine grained annotations of manipulation type, affected region or entity, reasoning category, and explanatory justification, designed to stress contextual grounding and explanation fidelity. Extensive experiments show that FakeHunter surpasses strong multimodal baselines, and ablation studies confirm that both contextual retrieval and selective tool activation are indispensable for improved robustness and explanatory precision.
\end{abstract}

\section{Introduction}

Recent advances in generative models~\cite{gao2023masked,peebles2023scalable,chen2024gaussianvton,chen2025hierarchical,chen2025temporal} and large language models~\cite{achiam2023gpt,liu2024deepseek,huang2025vistadpo,team2023gemini,chen2024uncertainty,chen2024bovila,du2025dependeval,li2025text} have dramatically lowered the barrier to creating realistic fake audio, images, and videos~\cite{tolosana2020deepfakes,hu2024recent,zheng2024videogen}. This widespread accessibility has sparked growing concerns: deepfakes have been exploited to spread political misinformation~\cite{chesney2019deepfakes,vaccari2020deepfakes}, conduct voice-cloning scams and impersonation attacks~\cite{korshunov2018deepfakes}, and undermine trustworthiness in video applications~\cite{mirsky2021creation,agarwal2020detecting,huang2024evidential,huang2024crest,huang2025trusted,chen2024finecliper,ma2024beyond,liu2023adaptive,liu2024adaptive}. Beyond simple facial manipulations, modern techniques enable audio spoofing~\cite{tak2021end}, object removal~\cite{mittal2023videosham}, and full-scene re-synthesis~\cite{bar2024lumiere}. These trends highlight the need for deepfake detection systems that are not only accurate, but also multimodal and explainable, capable of analyzing both visual and auditory signals to identify what was manipulated, where the manipulation occurred, and why it is considered fake.

Despite progress in deepfake detection, most existing methods remain limited in scope and capability. Some models are unimodal—operating solely on either visual~\cite{rossler2019faceforensics++,li2020celebdf,frank2020frequency} or audio~\cite{wang2023rawnet,jung2022aasist,di2025end} inputs—and fail to capture inconsistencies across modalities. Others incorporate both audio and visual streams~\cite{Yang2023AVoiDDF,Nie2024FRADE,Oorloff2024AVFF}, but remain confined to binary classification, offering no insight into what was manipulated or why the content is considered fake. 

To improve interpretability, recent work has introduced explanation-aware deepfake detectors. TAENet~\cite{du2024taenet} generates pixel-level heatmaps to localize manipulated regions but lacks textual reasoning. FakeShield~\cite{xu2024fakeshield} identifies segment-level audio-visual mismatches and produces human-understandable explanations based on visual-semantic inconsistencies. DD-VQA~\cite{zhang2024common} employs chain-of-thought reasoning via visual question answering prompts, while TruthLens~\cite{kundu2025truthlens} leverages vision and language models to generate textual justifications for face-image detection. However, all of these methods operate on static images or unimodal inputs, limiting their ability to reason over multimodal or dynamic manipulations. Thus, this highlights the need for a unified multimodal framework capable of both detection and explanation.

In this paper we present \mymodel as an explainable multimodal deepfake detection framework that integrates memory-guided retrieval, Chain-of-Thought (CoT) reasoning, and tool-augmented analysis. \mymodel encodes video and audio with pretrained CLIP and CLAP, retrieves semantically similar authentic samples from a FAISS-indexed memory bank, and employs them to ground step-by-step reasoning executed by Qwen2.5-VL. The pipeline follows an Observation–Thought–Action sequence to localize manipulations, articulate where and why they occur, and invoke zoom-in or spectrogram tools under uncertainty. To support systematic evaluation we construct \mydataset as a large-scale multimodal deepfake corpus containing over 5,700 videos (950+ minutes). Each instance involves either visual object removal or audio replacement and is annotated with the manipulated entity and timestamp, manipulation type, violated reasoning category, and a textual explanation. This resource enables rigorous assessment of both detection accuracy and interpretability in complex audio–visual settings. \mymodel achieves 34.75\% accuracy, surpassing state-of-the-art baselines. Our contributions are summarized as follows:

\begin{itemize}
    \item \textbf{Explainable and Multimodal Model.} We present \mymodel, a unified model that detects deepfakes and generates natural language explanations through memory-guided, step-by-step reasoning.
    \item \textbf{Multi-stage Reasoning Pipeline.} The model performs iterative Observation–Thought–Action steps with explanation verification for robust and interpretable results.
    \item \textbf{Fine-grained Multimodal Dataset.} We construct \mydataset, a new dataset of 5.7K+ videos (950+ minutes), featuring fine-grained annotations for visual object removal and audio replacement.
\end{itemize}

\section{Related Work}

\subsection{Deepfake Detection Methods}
Recent advances in deepfake detection span audio, visual, and multimodal domains. In audio, early methods relied on handcrafted features and shallow classifiers~\cite{QIAN201643}, while recent approaches use raw waveform CNNs~\cite{di2025end,tak2021end,wang2023rawnet}, spectro-temporal graph attention~\cite{jung2022aasist}, and semi-supervised GNNs with non-contrastive pretraining~\cite{febrinanto2025vision}. In visual, detectors target artifacts such as eye blinking~\cite{li2018ictu}, facial warping~\cite{rossler2019faceforensics++}, and frequency anomalies~\cite{frank2020frequency}. Though effective on known manipulations, unimodal models often degrade under compression or out-of-distribution attacks~\cite{li2020celebdf}. To enhance robustness, researchers explore data augmentation~\cite{agarwal2020detecting}, anomaly detection, and localized boundary modeling (e.g., Face X-ray~\cite{li2020facexray}). To address cross-modal inconsistencies, recent work fuses audio–visual signals: AVFF~\cite{Oorloff2024AVFF} learns joint features, FRADE~\cite{Nie2024FRADE} injects audio cues into vision transformers, and AVoiD-DF~\cite{Yang2023AVoiDDF} detects lip-speech misalignment via dual-stream encoder-decoder. Generalization is further improved through facial action units~\cite{Bai_2023_CVPR}, self-supervised audio-visual alignment~\cite{Feng_2023_CVPR}, and temporal style consistency~\cite{Choi_2024_CVPR}.

Despite these advances, most deepfake detectors are limited to binary classification—identifying whether content is manipulated—without explaining why it is fake or where the manipulation occurs. To improve interpretability, recent work leverages large language models (LLMs) and vision-language systems. FakeShield~\cite{xu2024fakeshield} and TruthLens~\cite{kundu2025truthlens} generate natural language rationales grounded in visual artifacts. Explanatory VQA frameworks like DD-VQA~\cite{zhang2024common} introduce chain-of-thought reasoning to justify decisions. Methods such as TAENet~\cite{du2024taenet} visualize manipulated regions via dual-decoder architectures, while Aghasanli et al.~\cite{aghasanli2023interpretable} use prototype-based interpretability to retrieve similar forgeries as supporting evidence. While promising, most explanation-aware approaches remain limited to static images or single modalities, motivating our framework for multimodal, reasoning-driven detection and explanation.

\subsection{Existing Datasets}
FaceForensics++~\cite{rossler2019faceforensics++} is one of the earliest large-scale benchmarks for facial manipulation, covering diverse forgery techniques. Celeb-DF~\cite{li2020celebdf} enhances video realism by reducing visual artifacts, while DFDC~\cite{dolhansky2019dfdc} scales up subject diversity for real-world evaluation. VideoSham~\cite{mittal2023videosham} broadens scope beyond facial forgeries to include professionally edited videos. DeeperForensics-1.0~\cite{jiang2020deeperforensics} provides 10K high-fidelity actor deepfakes with perturbations simulating real-world noise, and WildDeepfake~\cite{zi2020wilddeepfake} collects in-the-wild videos under unconstrained conditions. For multimodal evaluation, FakeAVCeleb~\cite{khalid2021fakeavceleb} fuses audio and visual manipulations to enable cross-modal consistency analysis. Recently, ExDDV~\cite{hondru2025exddv} introduced region-level annotations and textual rationales for explainable video deepfake detection. While these datasets have significantly advanced the field, each shows limitations in diversity, realism, or modality. Our benchmark aims to fill these gaps by providing more diverse, realistic, and multimodal deepfake data as a complementary resource.
\section{Dataset}


To enable explainable deepfake detection across both audio and visual modalities, we introduce \mydataset, a novel benchmark comprising dual-modality manipulations paired with grounded natural language reasoning. The dataset includes two primary types of tampering: (1) \textit{visual object removal} and (2) \textit{audio content replacement}. Each manipulated sample is generated with task-level instruction and annotated with fine-grained labels for manipulation region, type, and justification. In total, \mydataset~contains over 5,700 video sessions, spanning more than 950 minutes of content.

\begin{table}[htbp]
\centering
\begin{tabular}{cl}
\toprule
\textbf{Label} & \textbf{Category}  \\
\midrule
A & Physical Laws  \\
B & Time/Season  \\
C & Location/Culture   \\
D & Role/Profession   \\
E & Causality/Order \\
F & Narrative Context  \\
\bottomrule
\end{tabular}
\caption{7 Categories of Reasoning Violations}
\label{fig:dataset_a}
\end{table}

\subsection{Dataset Generation}

To ensure both realism and explainability, we follow a two-stage pipeline for generation: (1) identifying meaningful manipulation targets with explanations, and (2) executing tampering using modality-specific editing tools.

\subsubsection{Manipulation Target Identification.}
\begin{figure}[htbp]
\centering
\includegraphics[width=1\linewidth,height=0.22\textheight]{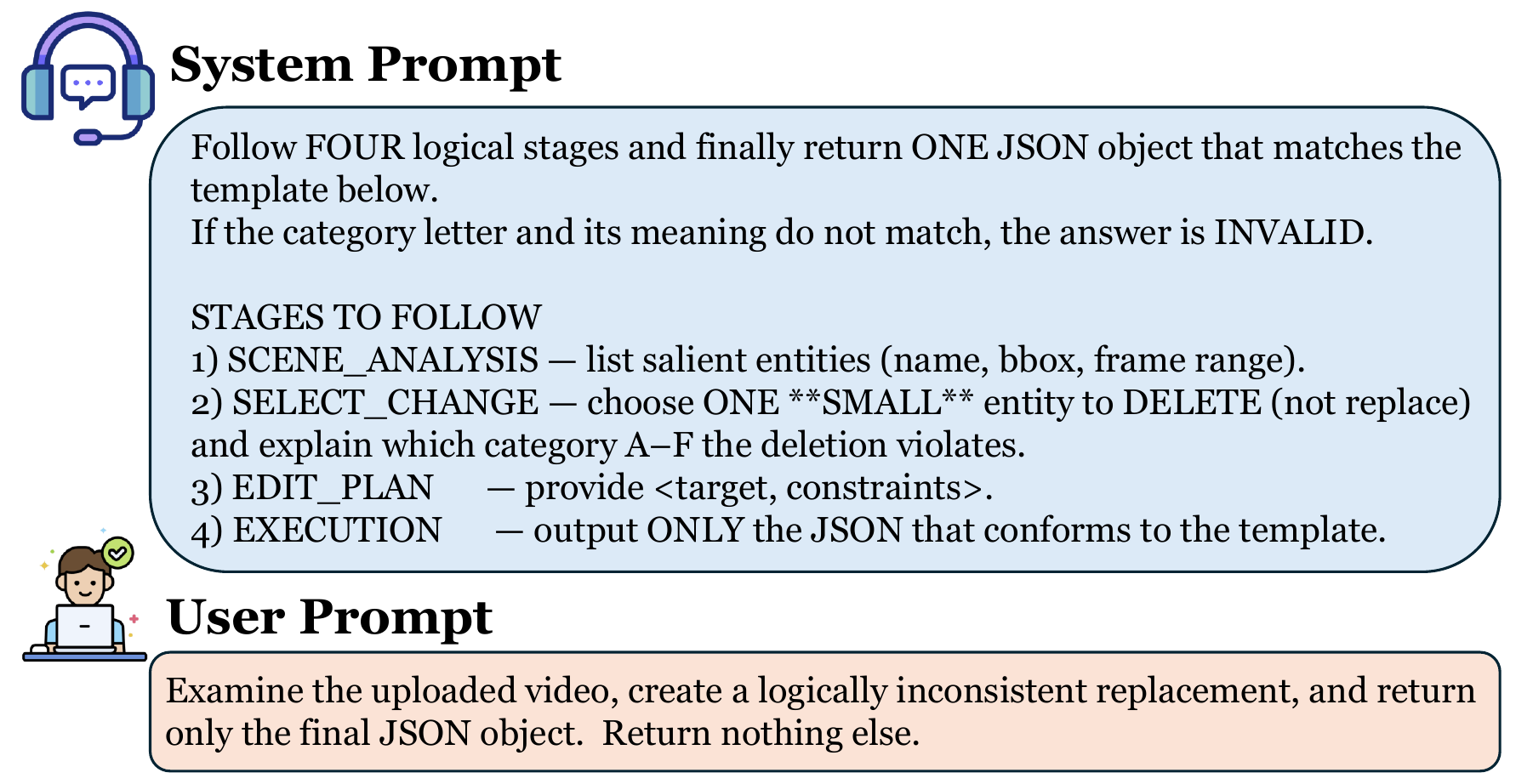}
\caption{Prompts that instruct VLM to do reasoning. We provide six choices about the logic of the violation, and after the VLM correctly understands the original video, it will follow the guidelines to select the appropriate object and give the reason with the logic of the violation}
\label{fig:dataset_b}
\end{figure}

\
\newline
We employ \texttt{Qwen2.5-VL}~\cite{bai2025qwen2} to analyze the input video and automatically determine (1) which entity to manipulate and (2) why such manipulation would introduce semantic or logical inconsistency. To guide Qwen's reasoning, we first define a taxonomy of seven reasoning violation types (labeled A–F), as illustrated in Figure~\ref{fig:dataset_a}. Then, we construct a structured system prompt (shown in Table~\ref{fig:dataset_b}) that asks the model to analyze the video/audio and output a manipulation plan in a standardized JSON format. This output includes the manipulation type (\texttt{visual-delete} or \texttt{audio-replace}), target entity, violated category, natural language justification, and temporal anchor (i.e., frame or timestamp). This structured JSON serves as the blueprint for grounded audio-visual manipulation.

\subsubsection{Manipulation Execution.}
\
\newline
Given the structured JSON plan, we apply specialized editing tools to manipulate either the video or audio content, depending on the specified modality and manipulation type.

\begin{itemize}
    \item \textbf{Video Manipulation}: We use \textit{Grounded SAM 2}~\cite{ren2024grounded} to track the target object and generate masks, then apply ProPainter~\cite{zhou2023propainter} to remove the object with spatial-temporal continuity.
    \item \textbf{Audio Manipulation}: We use \textit{Seeing-and-Hearing} model~\cite{xing2024seeing} to replace audio segments based on visual cues, producing mismatched speaker voices or out-of-sync sounds.
\end{itemize}

\begin{figure}[htbp]
\centering
\includegraphics[width=1\linewidth]{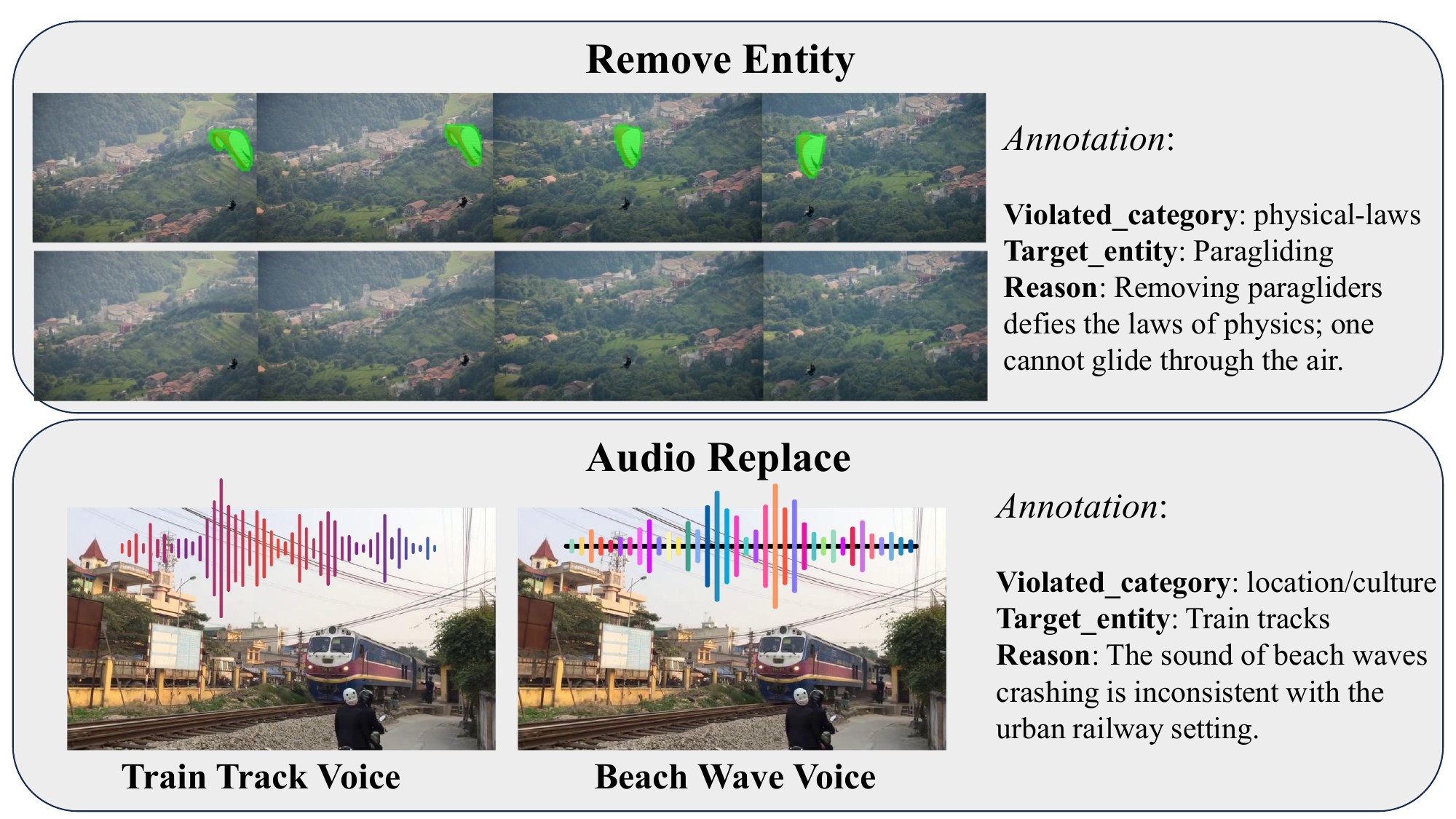}
\caption{\mymodel Dataset. For visual content, LLM will first select reasonable objects and give reasons for removing objects with violated rules; according to the guideline, we will call SAM2 for tracking and utilize tools for inpaint. For audio content, LLM analyzes the entire video and provides high-level guidance on how to perform semantically coherent audio replacement. Based on this guidance, we employ the audio toolkit to generate manipulated audio segments that align with the intended tampering strategy.}
\label{fig:dataset_example}
\end{figure}

\subsection{Dataset Overiew}

As illustrated in Figure~\ref{fig:dataset_example}, each sample in \mydataset~contains a pair of original and manipulated videos or audios, accompanied by structured metadata to support explainable detection and reasoning supervision. Specifically, we store:

\begin{itemize}
    \item \textit{Video/Audio IDs}: File ID of the original and tampered clips.
    \item \textit{Reasoning Trace}: A JSON object detailing manipulation type, target entity, violated category, explanation, and temporal anchor.
    \item \textit{Manipulation Annotations}: Pixel-level masks (visual) or timestamps (audio).
    \item \textit{Explanation Labels}: One of seven violation categories (A–F) and a natural language justification.
\end{itemize}

This rich annotation enables models trained on \mydataset~to move beyond binary classification and instead answer \textit{where}, \textit{what}, and \textit{why} content has been manipulated.

\section{Method}

\begin{figure*}[htbp]
\centering
\includegraphics[width=1\linewidth]{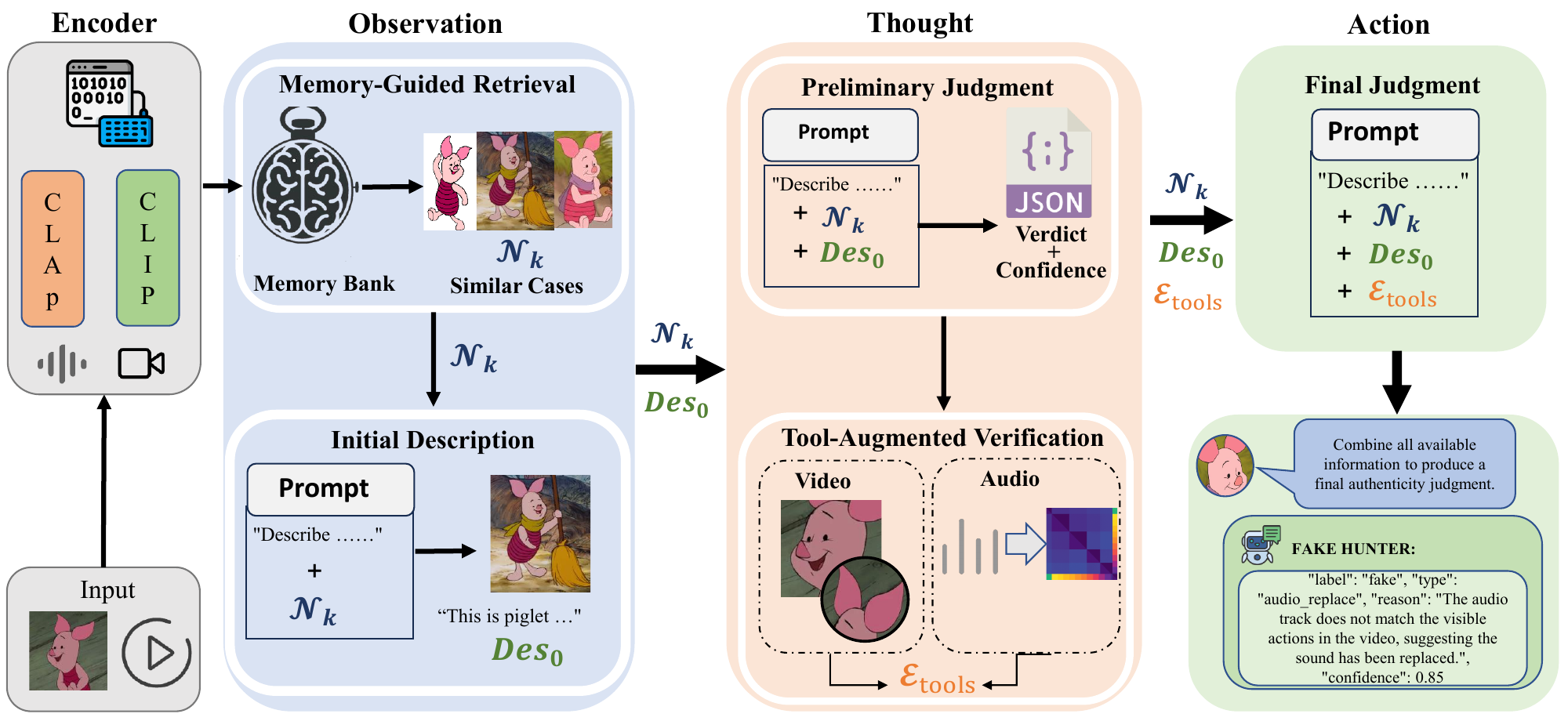}
\caption{Illustration of \mymodel. The framework follows an Observation–Thought–Action process. In the Observation stage, audio-visual features are encoded and used to retrieve semantically similar real examples from a memory bank, guiding the generation of an initial content description. In the Thought stage, the model makes a preliminary judgment based on the description and retrieved context; if confidence is low, visual and audio analysis tools are triggered for detailed inspection. In the Action stage, all available evidence is integrated to produce a final authenticity verdict with an interpretable explanation including manipulation type, location, and reasoning.}
\label{fig:overview}
\end{figure*}

We propose \mymodel, a benchmark for explainable multimodal deepfake detection. As illustrated in Figure~\ref{fig:overview}, our framework adopts a Chain-of-Thought (CoT) reasoning strategy that decomposes the decision-making process into an \textit{Observation–Thought–Action} sequence. By incorporating memory-guided retrieval and tool-augmented fine-grained analysis, \mymodel progressively improves both detection robustness and the quality of explanations.

\subsection{CoT Reasoning Pipeline}

To support explainable deepfake detection, we adopt a Chain-of-Thought (CoT) reasoning strategy that decomposes the decision-making process into an \textit{Observation–Thought–Action} sequence. This structured approach enables the model to reason over multimodal content in stages by integrating perception, comparison, and justification. Crucially, it allows the system not only to detect deepfakes but also to explain \textit{what}, \textit{where}, and \textit{why} a manipulation has occurred—thereby supporting transparent and interpretable video forensics.

\subsubsection{Observation.}

Let $(\mathcal{V}, \mathcal{A})$ denote the input video and audio streams, and $\mathcal{M}$ represent the memory bank. We first perform memory-guided retrieval to retrieve the top-$k$ most semantically similar exemplars $\{\mathbf{x}^{(i)}\}_{i=1}^k$ from the memory bank $\mathcal{M}$ using FAISS:
\begin{equation}
    \mathcal{N}_k = \mathrm{Search}(\mathbf{x}, \mathcal{M})
\end{equation}

Then, we leverage the retrieved set $\mathcal{N}_k$ as grounding context and prompt the VLM using a structured description prompt to produce an initial high-level interpretation of the input content:
\begin{equation}
    \mathit{Des} = \mathrm{VLM}(\mathcal{V}, \mathcal{A}, \mathcal{N}_k, \mathit{P}_{0})
\end{equation}

\subsubsection{Thought.}

Given the initial description $\mathit{Des}$ and the retrieved memory context $\mathcal{N}_k$, we prompt a VLM to form a preliminary judgment:

\begin{equation}
\mathcal{V} \triangleq \mathrm{VLM}(\mathit{Des},\ \mathcal{N}_k,\ \mathit{P}_{1})
\end{equation}

The model output contains four fields:
\[
\mathcal{V} \triangleq \{\mathit{L},\ \mathit{T},\ \mathit{E},\ \mathit{C}\},
\]
where $\mathit{L}$ indicates the binary classification result ($\mathit{real}$ or $\mathit{fake}$); $\mathit{T}$ specifies the manipulation modality (e.g., $\mathit{audio\text{-}replace}$, $\mathit{visual\text{-}delete}$); $\mathit{E}$ provides a natural language explanation grounded in the multimodal context; and $\mathit{C}$ denotes the model’s self-assessed prediction certainty, scaled to the range $[0, 1]$.

If the confidence score falls below a predefined threshold $\tau$, we invoke the Tool-Augmented Verification module to perform additional fine-grained analysis.

\subsubsection{Action.}

In the final stage, the model consolidates all available information, including the descriptive summary $\text{Des}_0$, retrieved exemplars $\mathcal{N}_k$, and optional tool-based analysis $\mathcal{E}_{\text{tools}} $, into a final decision. We re-prompt a VLM to generate the final result $\mathcal{R}$:
\begin{equation}
\mathcal{R} \triangleq \mathrm{VLM}(\mathit{Des}, \mathcal{N}_k, \mathcal{E}_{\text{tools}} , \mathit{P}_{2})
\end{equation}

The final decision includes the following fields:
\[
\begin{aligned}
\mathcal{R} \triangleq \{\, &\mathit{L},\ \mathit{T},\ \mathit{E} \,\}
\end{aligned}
\]
where $\mathit{L}$ specifies authenticity, $\mathit{T}$ denotes manipulation type, and $\mathit{E}$ is a concise, human-readable justification of why the content is considered fake.

\subsection{Memory-Guided Retrieval}

\subsubsection{Feature Encoding}
\
\newline
We extract modality-specific features from each input video to form a joint audio-visual representation:

\begin{itemize}
    \item \textbf{Visual Features.} We uniformly sample $T$ keyframes $\{I_t\}_{t=1}^{T}$ and encode each frame using a pretrained CLIP image encoder~\cite{radford2021learning}, yielding image features $f^{\text{img}}_t \in \mathbb{R}^{d_1}$.
    
    \item \textbf{Audio Features.} The corresponding audio track is segmented into $T$ chunks $\{A_t\}_{t=1}^{T}$ aligned with each keyframe. We apply the CLAP audio encoder~\cite{elizalde2023clap} to obtain $f^{\text{aud}}_t \in \mathbb{R}^{d_2}$.
\end{itemize}

We concatenate the two modalities to obtain a fused multimodal feature:
\begin{equation}
    f_t = [f^{\text{img}}_t \, \| \, f^{\text{aud}}_t],~~ f_t \in \mathbb{R}^{d}.
\end{equation}

We average the per-segment embeddings to obtain a video-level representation:
\begin{equation}
    \mathbf{f}_v = \frac{1}{T} \sum_{t=1}^{T} f_t,~~  \mathbf{f}_v \in \mathbb{R}^{d}.
\end{equation}

\subsubsection{Memory Bank Construction and Retrieval}
\
\newline
To more reliably detect subtle manipulations, we construct a memory bank $\mathcal{M}$ of real videos and retrieve semantically similar exemplars at inference time. These authentic references serve as the grounding context, enabling the model to contrast the input against typical audio-visual patterns and more effectively identify anomalies.

We perform K-Means clustering over training set embeddings $\{\mathbf{f}_v\}$ to select $K=300$ cluster centers as representative memory anchors. These prototypes are indexed using FAISS~\cite{johnson2019billion} to support efficient similarity-based retrieval at inference.

At test time, given an input video representation $\mathbf{f}_q$, we retrieve its top-$k$ nearest neighbors from the memory bank:
\begin{equation}
    \mathcal{N}_k(\mathbf{f}_q) = \underset{\mathbf{f} \in \mathcal{M}}{\text{Top-}k}\left( \frac{\mathbf{f}_q^\top \mathbf{f}}{\|\mathbf{f}_q\| \cdot \|\mathbf{f}\|} \right).
\end{equation}

The retrieved set $\mathcal{N}_k$ is integrated into the reasoning process as contextual support, enabling comparative inference for detecting inconsistencies and producing explanations.

\subsection{Tool-Augmented Verification}
To enhance robustness under low-confidence predictions, we introduce a conditional tool-based verification stage that is invoked only when necessary. Let $ \gamma \in [0,1] $ denote the model confidence for the current sample, and fix the activation threshold at $ \tau = 0.80 $. The binary trigger is defined as
\begin{equation}
\mathrm{Trigger}(\gamma) = \mathbb{1}\{\gamma < \tau\},
\end{equation}
which evaluates to $1$ precisely in the low-confidence regime ($ \gamma < \tau $) and $0$ otherwise. This selective mechanism reduces unnecessary computational overhead while improving reliability in uncertain regions.

\subsubsection{Visual Zoom-In Analysis.}
\
\newline
Let $\mathcal{I}_0$ be the reference frame from video $\mathcal{V}$, and let $\mathcal{R} \subset \mathcal{I}_0$ denote a cropped region of interest (ROI) suspected of containing visual artifacts or tampering evidence. This region is then passed to a vision-language model for localized multimodal reasoning. And the model returns a descriptive report highlighting local visual anomalies such as edge seams, spatial distortions, or inconsistent lighting.
\subsubsection{Audio Spectrogram Analysis.}
\
\newline
If audio is present, we temporally align the suspect region with audio segment $\mathcal{A}_t$ and compute its mel-spectrogram:
\begin{equation}
\mathcal{S}_t = \mathcal{G}(\mathcal{A}_t) \in \mathbb{R}^{F \times T}
\end{equation}
where $F$ and $T$ denote frequency bins and time steps, respectively. And $\mathcal{G}$ denotes the function of computing mel-spectrogram The resulting spectrogram $\mathcal{S}_t$, encoding temporal and spectral features, is then analyzed by a VLM.
The outputs from these modules collectively form a structured evidence tuple containing multimodal justifications and localization cues:
\begin{equation}
\mathcal{E}_{\text{tools}} \triangleq \{V(\mathcal{R}), A(\mathcal{S}_t)\},
\end{equation}
where $V(\mathcal{R})$ and $A(\mathcal{S}_t)$ represent the video and audio report from VLM.
The combined evidence is integrated into the final stage of CoT reasoning to refine the authenticity judgment and produce detailed, interpretable explanations.

\section{Experiments}

\begin{table*}[ht]
\centering
\resizebox{\textwidth}{!}{ 
\begin{tabular}{lcccccccc}
\toprule
\textbf{Dataset} & \textbf{Year} & \textbf{Modality} & \textbf{Application} & \textbf{Manipulated} & \textbf{\# Attacks} & \textbf{\# Real} & \textbf{\# Fake}  & \textbf{Explanation}\\
\midrule
MTVFD & 2016  & V & Video Manipulation & User Generated & 1 & 30 & 30 & x\\
UADFV & 2018  & V & Face & Deep Learning& 3 & 49 & 49 & x\\
FaceForensics++& 2019  & V & Face & Deep Learning & 4 & 1000 & 4000 & x\\
CelebDF& 2020 & V & Face & Deep Learning & 3 & 5907 & 5639 & x\\
WildDeepFake& 2021  & V & Face & Deep Learning & 4 & 3805 & 3509 & x\\
Psynd & 2022  & A & Speech & Deep Learning & 1 & 30 & 2,371 & x\\
VideoSham& 2022  & A+V & Video Manipulation & User Generated & 6 & 380 & 380 & x\\
FakeBench& 2024 & I & Image Manipulation& Deep Learning & 6 & 3000 & 3000 &  \checkmark \\
VANE-Bench& 2024 & V & Video Manipulation & Deep Learning & 5 & 1000 & 2000 &  \checkmark \\
\mydataset (Ours) & 2025  & A+V & Video Manipulation & Deep Learning & 2 & 5700 & 5700 & \checkmark \\
\bottomrule
\end{tabular}}
\vspace{-1em}
\caption{Comparison of datasets for deepfake and manipulation detection.}
\label{tab:dataset_comparison}
\end{table*}

\subsection{Experimental Setting}

\noindent \textbf{Implementation Details.}  
All experiments are conducted in PyTorch on a Linux server with 4$\times$NVIDIA A800 GPUs (80GB). Below we summarize key configurations:

\begin{itemize}
    \item \textbf{LLM Inference:} We use Qwen2.5-Omni-7B with bfloat16 precision and Flash Attention 2 enabled for efficient and memory-optimized inference. The model supports a maximum context length of 32,768 tokens, enabling long-range, coherent multimodal reasoning across video and audio inputs.
    
    \item \textbf{Video Preprocessing:} Each input video is sampled at 1 FPS and limited to 128 frames or 30 seconds of audio. We use a \texttt{medium} visual resolution setting for efficiency-accuracy trade-off.
    
    \item \textbf{Feature Embedding:} We extract 512-dimensional embeddings from CLIP (visual) and CLAP (audio), and concatenate them into a unified 1024-dimensional multimodal feature vector for downstream reasoning.
    
    \item \textbf{Memory Retrieval:} We retrieve the top-$k=5$ nearest neighbors from a FAISS-indexed memory bank containing up to 10,000 reference samples. A similarity threshold of 0.7 is applied for filtering.
    
    \item \textbf{Reasoning Configuration:} Each input is processed in up to 3 reasoning rounds. If the prediction confidence from Qwen2.5-Omni-7B falls below 0.8, tool-augmented verification is triggered.
\end{itemize}

\noindent \textbf{Baselines.}  
We evaluate both \mymodel{} and \mydataset{} against strong baselines. For dataset comparison, we benchmark \mydataset{} against 9 widely used datasets: MTVFD~\cite{al2016development}, UADFV~\cite{yang2019exposing}, FaceForensics++~\cite{rossler2019faceforensics++}, CelebDF~\cite{li2020celeb}, WildDeepFake~\cite{zi2020wilddeepfake}, Psynd~\cite{zhang2022localizing}, VideoSham~\cite{mittal2023video}, FakeBench~\cite{li2024fakebench}, and VANE-Bench~\cite{gani2025vane}.

For comparison, we consider two groups of baselines:
\begin{itemize}
    \item \textbf{LLM-based Reasoners:} Qwen2.5-Omni-7B and MiniCPM-o-2\_6.
    \item \textbf{Deepfake Detection Models:} We include strong modality-specific baselines such as FTCN~\cite{Oorloff2024AVFF} for audio-visual forgery classification and AASIST~\cite{jung2022aasist} for audio spoofing detection.

\item \textbf{Deepfake Reasoning Models:} Due to the lack of publicly available reasoning-based deepfake detectors, no direct comparison is made in this category.

\end{itemize}

\noindent \textbf{Dataset.}  We evaluate our method exclusively on \mydataset{}, as no existing dataset provides fine-grained annotations for both visual and audio manipulations along with corresponding reasoning labels. 

\subsection{Comparison to Competitive Dataset}

Table~\ref{tab:dataset_comparison} summarizes a comparison between \mydataset{} and nine representative deepfake or manipulation datasets across vision, audio, and multimodal domains. While prior datasets such as FaceForensics++~\cite{rossler2019faceforensics++}, CelebDF~\cite{li2020celeb}, and UADFV~\cite{yang2019exposing} focus primarily on facial visual forgeries, they lack audio content and fine-grained reasoning annotations. Similarly, audio-specific datasets like Psynd~\cite{zhang2022localizing} are unimodal and limited in manipulation types.

VideoSham~\cite{mittal2023video} and VANE-Bench~\cite{gani2025vane} include multimodal manipulations, but do not provide step-by-step explanations or well-defined reasoning categories for interpretability. FakeBench~\cite{li2024fakebench} is image-based and includes textual explanation, but lacks temporal dynamics and cross-modal complexity.

In contrast, \mydataset{} introduces both audio and video manipulations in the same benchmark, paired with structured metadata such as violated reasoning category (A–F), manipulated region/timestamp, and natural language justifications. It is the only dataset to support explainable multimodal deepfake detection with over 5,700 high-quality samples and balanced real/fake pairs, offering a unique testbed for detection and interpretability research.


\begin{table*}[ht]
\centering
\renewcommand{\arraystretch}{1.1}
\begin{tabular}{l|ccc}
\toprule
\textbf{Model} & \textbf{Overall Accuracy} & \textbf{Audio Subset} & \textbf{Visual Subset} \\
\midrule
FTCN (video only)      & —        & —         & 0.00\% \\
AASIST (audio only)    & —        & 18.69\%   & —      \\
Qwen2.5-Omni-7B        & 18.68\%  & 12.27\%   & 24.83\% \\
MiniCPM-o-2\_6         & 9.19\%   & 17.88\%   & 0.78\% \\
\mymodel{} with Qwen2.5-Omni-7B     & \textbf{34.75\%} & \textbf{23.00\%} & \textbf{46.50\%} \\
\mymodel{} with MiniCPM-o-2\_6     & 27.00\%  & 25.50\%   & 28.50\% \\
\bottomrule
\end{tabular}
\vspace{-1em}
\caption{Accuracy comparison across supported modalities. Multimodal models are evaluated on both audio and visual subsets.}
\label{tab:model_comparison}
\end{table*}

\subsection{Comparison to State-of-the-Art Approaches}

We compare \mymodel{} against a set of representative baselines, covering both unimodal and multimodal deepfake detection methods. The competing methods are categorized as follows:

\begin{itemize}
    \item \textbf{Unimodal Detection Models:} FTCN~\cite{Oorloff2024AVFF} for visual-only detection, and AASIST~\cite{jung2022aasist} for audio-only detection.
    \item \textbf{Multimodal LLM-Based Models:} Qwen2.5-Omni-7B and MiniCPM-o-2\_6, which accept both audio and visual inputs.
\end{itemize}

Unimodal baselines are evaluated solely on their supported modality, whereas \mymodel{} and LLM-based models are evaluated on the full multimodal dataset and modality-specific subsets. Specifically, we report:

\begin{itemize}
    \item \textbf{Audio Accuracy} over $\mathcal{D}_A$ (audio-manipulated videos),
    \item \textbf{Visual Accuracy} over $\mathcal{D}_V$ (visually-manipulated videos),
    \item \textbf{Overall Accuracy} over $\mathcal{D}_A \cup \mathcal{D}_V$:
    \[
    \text{Acc}_\text{Overall} = \frac{|\text{Correct}_A| + |\text{Correct}_V|}{|\mathcal{D}_A| + |\mathcal{D}_V|}.
    \]
\end{itemize}

Table~\ref{tab:model_comparison} presents the results. FTCN and AASIST, limited to single modalities, show suboptimal performance compared to multimodal approaches. Qwen2.5-Omni and MiniCPM-o-2\_6 demonstrate moderate baseline capability but struggle with multimodal alignment.

\mymodel{} achieves superior results across all metrics:

\begin{itemize}
    \item \textbf{Multimodal Advantage:} \mymodel(Qwen2.5) achieves 34.75\% on $\mathcal{D}_{\text{all}}$, nearly doubling Qwen2.5-Omni (18.68\%) and MiniCPM-o-2\_6 (9.19\%).
    \item \textbf{Enhanced Visual Detection:} Zoom-in visual inspection enables \mymodel(Qwen2.5) to reach 46.50\% accuracy on $\mathcal{D}_{\text{visual}}$.
    \item \textbf{Framework Gain:} While MiniCPM-o-2\_6 alone performs poorly (e.g., 0.78\% on visual), our framework boosts it to 28.50\% via retrieval, reasoning, and tool-based refinement.
\end{itemize}

These results highlight the effectiveness of our memory-guided CoT reasoning pipeline, demonstrating its capability to handle challenging multimodal deepfakes with strong interpretability and generalizability.

\begin{table*}[htbp]
\centering
\begin{tabular}{l|ccc}
\toprule
\textbf{Model Variant} & \textbf{Overall Accuracy} & \textbf{Audio Subset} & \textbf{Visual Subset} \\
\midrule
Qwen2.5-Omni-7B (Raw) & 18.68\% & 12.27\% & 24.83\% \\
\mymodel{} (Qwen) w/ Memory, w/o Tool & 21.00\% & 1.50\% & 40.50\% \\
\mymodel{} (Qwen) w/o Memory, w/ Tool & 27.00\% & 27.00\% & 27.00\% \\
\mymodel{} (Qwen) Full (w/ Memory \& Tool) & \textbf{34.75\%} & \textbf{23.00\%} & \textbf{46.50\%} \\
\bottomrule
\end{tabular}
\vspace{-1em}
\caption{Performance comparison of \mymodel{} under different ablation settings on the X-AVFake benchmark. Memory-guided retrieval and tool-augmented verification both contribute to improved accuracy.}
\label{tab:model_comparison}
\end{table*}

\subsection{Ablation Study}
To assess the contribution of each module in \mymodel{}, we perform an ablation study comparing four system variants on the X-AVFake benchmark (Table~\ref{tab:model_comparison}). We isolate the effects of \textit{Memory-Guided Retrieval} and \textit{Tool-Augmented Verification}.

\textbf{Base LLM Performance.} Without our framework, the raw Qwen2.5-Omni-7B achieves limited accuracy (18.68\% overall), with lower results on audio (12.27\%) than visual (24.83\%). This highlights the difficulty of detecting multimodal inconsistencies via end-to-end LLM inference alone.

\textbf{Effect of Tool-Augmented Verification.} Adding our tool module (e.g., zoom-in or spectrogram inspection) improves audio deepfake detection. Compared to the memory-only variant (21.00\%), removing memory while retaining tools yields 27.00\% accuracy—showing that localized, fine-grained analysis recovers subtle cues missed by global reasoning.

\textbf{Effect of Memory-Guided Retrieval.} The memory-only variant performs well on visual samples (40.50\%) but poorly on audio (1.50\%), suggesting visual grounding benefits more from retrieval contrast, while audio inconsistencies require deeper inspection.

\textbf{Full Model.} Our complete \mymodel{} (with memory and tools) outperforms all variants, reaching 34.75\% overall, with strong results across modalities (23.00\% audio, 46.50\% visual). This confirms that both contextual retrieval and fine-grained inspection are essential for robust, interpretable multimodal detection.

\subsection{Runtime Efficiency}

The full FakeHunter pipeline—including feature extraction, memory retrieval, and multimodal reasoning—runs at ~0.9$\times$ real-time on four NVIDIA A800 GPUs. Despite its multi-stage design with modular reasoning and tool-based inspection, inference remains practical. This efficiency stems from the lightweight Qwen2.5-Omni backbone and FAISS-accelerated memory retrieval.

\section{Conclusion}

\textbf{FakeHunter} is a step-by-step multimodal framework that integrates
\emph{memory retrieval}, \emph{Chain-of-Thought reasoning}, and \emph{tool-augmented verification}
for accurate \emph{and} interpretable deepfake detection.
Built on \textbf{Qwen2.5-Omni-7B}, it jointly analyzes video \emph{and} audio, anchors
judgments with retrieved exemplars, and invokes zoom-in or spectrogram tools
when confidence is low.
Each stage emits structured evidence describing \emph{what} was manipulated,
\emph{where} it occurred, and \emph{why} it is fake—bridging the gap between
raw predictions and forensics-grade explanations.

To support explainable forensics, we release \textbf{X-AVFake},  
a $\sim$5.7k-video benchmark annotated with manipulation type, target entity,
reasoning category, and free-form justification.
Covering visual entity removal and audio replacement, X-AVFake enables
holistic evaluation of accuracy \textit{and} interpretability across complex
audio–visual manipulations.

Experiments show FakeHunter achieves \textbf{34.75\%} accuracy,
surpassing vanilla Qwen2.5-Omni-7B by +16.9 pp and MiniCPM-2.6 by +25.6 pp.
Memory retrieval yields steady gains, while tool-based inspection is
\emph{crucial} for low-confidence segments, lifting accuracy to 46.5\%.
Despite its multi-stage design, the pipeline runs at
$\sim$0.3$\times$ real-time on 4$\times$NVIDIA A800 GPUs, processing a 10-min clip in
~33 minutes—demonstrating deployment feasibility.

\paragraph{Broader Impact.}
Explainable detection benefits journalism, moderation, and legal forensics by
revealing model reasoning and enabling court-admissible analysis.
Its modular design allows integration of future detectors or
domain-specific tools without retraining the core model.

\paragraph{Future Work.}
We plan to (1) expand X-AVFake with higher-order manipulations like scene
re-synthesis and cross-modal mismatches, (2) develop adaptive tool-selection
for faster inference, and (3) explore end-to-end training that
fuses reasoning traces with audio-visual features.
We will release dataset, code, and models to foster transparent,
community-driven progress in explainable video forensics.

\bibliography{aaai2026}

\end{document}